# Leveraging Machine Learning Techniques for Windows Ransomware Network Traffic Detection

Omar M. K. Alhawi, James Baldwin, Ali Dehghantanha

Department of Computer Science, School of Computing, Science and Engineering, University of Salford

O.Alhawi@edu.salford.ac.uk, J.Baldwin1@edu.salford.ac.uk, A.Dehghantanha@salford.ac.uk

**Abstract**

Ransomware has become a significant global threat with the ransomware-as-a-service model enabling easy availability and deployment, and the potential for high revenues creating a viable criminal business model. Individuals, private companies or public service providers e.g. healthcare or utilities companies can all become victims of ransomware attacks and consequently suffer severe disruption and financial loss. Although machine learning algorithms are already being used to detect ransomware, variants are being developed to specifically evade detection when using dynamic machine learning techniques. In this paper we introduce *NetConverse*, a machine learning analysis of Windows ransomware network traffic to achieve a high, consistent detection rate. Using a dataset created from conversation-based network traffic features we achieved a true positive detection rate of 97.1% using the Decision Tree (J48) classifier.

**Keywords**: Ransomware, Malware detection, Machine learning, Network traffic, Intrusion detection

# 1. Introduction

Ransomware has become a significant global threat in the last 2 years with the FBI estimating that $1Billion of ransom demands were paid in 2016; this represents a 400% increase from the previous year [1]. In the same period the U.S. experienced a 300% increase in the number of daily ransomware attacks [2] and the cost of the average ransom demand doubled [3]. At the end of 2015 Symantec logged a record number (100) of new ransomware families [3]. The increase over the 2-year period is attributed to the rise of ransomware-as-a-service (RaaS) model. RaaS provided the cybercriminal with the ability to purchase ransomware creation kits and source code and distribute ransomware with very little technical knowledge [4].

In 2016 Europol reported that ransomware had become the primary concern for European law enforcement agencies with Cryptoware (the class of ransomware that encrypts files in comparison to the decreasingly prevalent locker class) the most prominent malware threat [5]. In July 2016 the No More Ransom project [6] was launched as a partnership between European law enforcement and IT Security companies in an attempt to disrupt ransomware related criminal activities, and help businesses and individuals mitigate against the impact of ransomware. Similarly commercial software products have been developed to defend networks; Cybereason [7] uses behavioural techniques to protect consumer networks; Darktrace [8] has employed advanced unsupervised machine learning to protect enterprise networks.

Several machine learning techniques and frameworks have been proposed and undertaken for ransomware detection. However, dynamic analysis techniques have limitations in that new ransomware variants can be redesigned in an attempt to decrease the rate of detection by machine learning algorithms [9]. The application of machine learning for dynamic analysis of ransomware has achieved detection rates >96% [10]. Similarly, the application of machine learning for network traffic analysis of Android malware has achieved detection rates >99% [11].

The vast majority of ransomware threats today are designed to target personal computers running the Windows operating system since Windows-based computers make up around 89 percent the OS market share of desktop computers [12]. The NetConverse model will focus on the Windows environment and proposes to leverage machine learning techniques for detecting Windows ransomware through analysing network traffic. The contributions made to achieve this goal are as follows:

(a) The evaluation study analysed network traffic for 9 ransomware families with 210 samples, and 3 goodware types with 264 samples.
(b) 6 machine learning classifiers were evaluated and classified into 3 groups.
(c) 13 features from ransomware samples traffic were extracted and deployed. These samples were extracted using TShark [13].

The paper is organised as follows: Section 2 discusses related works on the topic of machine learning, ransomware detection and network traffic analysis; Section 3 describes the research methodology and the three phases that it comprises; Section 4 presents the experiment and discusses the results; Section 5 presents the conclusion and discusses future works.

# 2. Related Works

Malicious programs and exploit kits have always been important tools in cyber criminals' toolset [14], and machine learning techniques have been used for decades for malware detection and analysis [15].

Different detection, analysis and investigation approaches have been proposed to defend against malware however malicious programs are employing a variety of propagation and evasion techniques to bypass defensive mechanisms [16]. Malware classification using machine learning has proved very successful in the detection of Android malware [17]. Malware behavioural characteristics such as API calls, filesystem changes and network traffic have been used as features for different classification tasks [18], hence machine learning can be utilised for Ransomware detection as well [19].

Ransomware can be categorized into two main classes: Locker ransomware denies access to the computer or device [12]; Crypto ransomware prevents access to files or data. As ransomware samples are using different evasion techniques, any ransomware analysis should take these techniques into account [20], [21]. Previous works for ransomware detection and analysis can be divided into static and dynamic approaches [22]; static approaches are relying on ransomware signature or utilisation of a cryptographic primitive function for detection [23]; dynamic methods are using dynamic binary instrumentation such as ransomware runtime activities for detection [24].

EldeRan [10], which was a ransomware classifier based on a sample's dynamic features, could achieve a True Positive Rate (TPR) of 96.3% with a low False Positive Rate (FPR) of 1.6%. UNVEIL [25] is another machine learning based system for ransomware detection using a ransomware sample interacting with the underlying O.S. which achieved a True Positive Rate (TPR) of 96.3%, and a zero False Positive Rate (FPR). Network behaviours and Netflow data can also be used for ransomware detection [26].

With NetConverse we are investigating the application of different machine learning classifiers in detecting ransomware samples using features extracted from network traffic conversations; we report accuracy, TPR and FPR as metrics to evaluate the performance of our classification tasks.

## 3. Methodology

This section presents the data collection, feature extraction and machine learning classifier phases of our experiment. The 3 phases are outlined in Fig. 1.

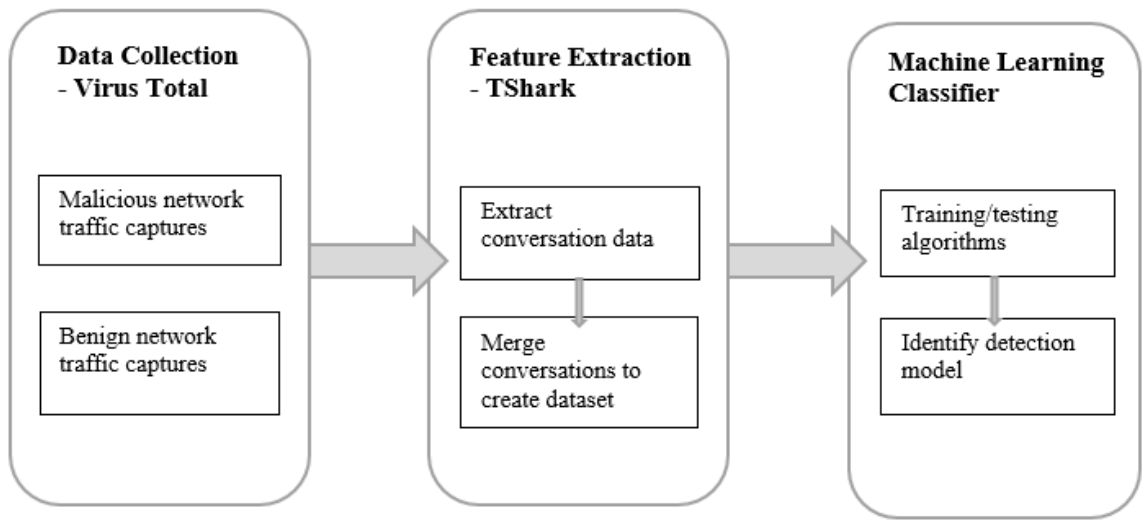

**Fig. 1** Workflow showing the 3 phases for the experiment

In the data collection phase, the network traffic samples are collected for both malicious (ransomware) and benign Windows applications. The feature extraction phase extracts the relevant features and merges them to create our dataset. In the final machine learning classifier stage, we train and test several algorithms in the Waikato Environment for Knowledge Analysis 3.8.1 (WEKA) machine learning tool [27] to identify the optimum detection model.

3.1 Data collection phase

In this experiment, we are focusing on Windows ransomware network traffic and looking at the characteristics of the network conversations created when a host is infected. The infected host will attempt to connect to a remote attacker network address which could be a command and control server, payment or distribution website [28]. Thus, these are the network conversations that we are capturing and comparing with the characteristics of a benign applications using different classification techniques.

*3.1.1 Malicious applications*

We identified the ransomware families to be included in our experiment by looking at the current tracked families on the Ransomware Tracker website, namely Cerber, Cryptowall, CTB-Locker, Locky, Padcrypt, Paycrypt, Teslacrypt and Torrentlocker. The site provides an overview of the internet infrastructure used by ransomware cyber criminals [28]. Cryptolocker was also included due to its prevalence prior to 2014. By searching for the specific ransomware families in the Virus Total Intelligence platform [29] we could identify portable executable samples that had a corresponding behavioural analysis network traffic capture. For each sample the PCAP file was downloaded and saved with the name of the sample hash value. In total, we collected 210 network traffic captures which are summarised in Table 1.

**Table 1**
Summary of ransomware families

| Ransomware Family | Class  | Sample size |
|-------------------|--------|-------------|
| Cerber            | Crypto | 30          |
| Cryptowall        | Crypto | 30          |
| Cryptolocker      | Crypto | 30          |
| CTB-Locker        | Locker | 30          |
| Locky             | Locker | 30          |
| Padcrypt          | Crypto | 1           |
| Paycrypt          | Crypto | 2           |
| Teslacrypt        | Crypto | 27          |
| Torrentlocker     | Locker | 30          |

*3.1.2 Benign applications*

Our benign network traffic samples (goodware) were also collected from the Virus Total Intelligence platform [29]. Our search criteria targeted portable executable files that had been submitted at least 3 times and had 0 detections by antivirus engines. The search was applied to several different behavioural report criteria to provide a collection of 264 goodware samples.

3.2 Feature selection and extraction:

Feature extraction was achieved using TShark, a terminal based feature of the network protocol analyser Wireshark [30]. The network traffic capture PCAP files can be analysed within Wireshark, but it offers limited export features. TShark provides a more flexible, powerful export feature that can create statistical, calculated data in addition to static feature extraction. We chose to extract several basic, traffic and connection based features using the TShark conversations export option. This aggregated each network capture file into unique conversations based on the 5-tuple [31] protocol, source/destination IP address, source/destination port values; the equivalent statistical values were also extracted. Each export file was merged together to create our initial pre-processed data. Table 2 shows the list of extracted features. Table 3 shows a sample of the dataset before cleaning.

**Table 2**
List of extracted features

| Feature name | Type [32], [33] | Description |
| --- | --- | --- |
| Protocol | Basic | Protocol type |
| Address A | Basic | Source IP address |
| Port A | Basic | Source host port number |
| Address B | Basic | Destination IP address |
| Port B | Basic | Destination host port number |
| Packets | Basic | Total number of packets per conversation |
| Bytes | Basic | Total number of bytes per conversation |
| Packets A → B | Connection | The number of packets from source to destination |
| Bytes A → B | Connection | The number of bytes sent from source to destination |
| Packets B → A | Connection | The number of packets sent from destination to source |
| Bytes B → A | Connection | The number of bytes sent from destination to source |
| Rel Start | Basic | Time relative to start of the conversation (seconds) |
| Duration | Basic | Duration of the conversation (seconds) |

**Table 3**
Dataset sample

| Protocol | Address A | Port A | Address B | Port B | Packets | Bytes | Packets A → B | Bytes A → B | Packets B → A | Bytes B → A | Rel Start | Duration |
| --- | --- | --- | --- | --- | --- | --- | --- | --- | --- | --- | --- | --- |
| 17 | 192.168.56.15 | 59612 | 91.121.216.96 | 6892 | 0 | 0 | 1 | 56 | 1 | 56 | 73.18 | 0 |
| 6 | 192.168.56.17 | 58762 | 2.18.213.64 | 80 | 124 | 164942 | 64 | 15590 | 188 | 180532 | 92.43 | 118.81 |
| 6 | 192.168.56.26 | 57258 | 216.58.214.78 | 443 | 2699 | 3935279 | 1107 | 100421 | 3806 | 4035700 | 151.63 | 117.63 |
| 6 | 192.168.56.19 | 55909 | 172.217.16.174 | 443 | 42 | 34349 | 33 | 11327 | 75 | 45676 | 176.14 | 116.65 |
| 6 | 192.168.56.17 | 58768 | 23.37.54.100 | 80 | 4 | 1537 | 7 | 1302 | 11 | 2839 | 93.60 | 107.63 |
| 6 | 192.168.56.14 | 49278 | 216.58.214.78 | 443 | 2694 | 3934720 | 610 | 62847 | 3304 | 3997567 | 184.56 | 104.34 |
| 6 | 192.168.56.14 | 49273 | 35.161.88.115 | 443 | 13 | 4509 | 15 | 1642 | 28 | 6151 | 183.90 | 61.66 |
| 6 | 192.168.56.17 | 58767 | 54.225.100.50 | 80 | 4 | 447 | 6 | 1070 | 10 | 1517 | 93.32 | 61.61 |

The pre-processed data was analysed to remove any outliers and erroneous records. We removed records with an Address A value of 0.0.0.0 and Port B records with a value of 53 which represented DNS traffic. The Packets, Bytes, Rel.Start and Duration attributes were also removed to leave us with 9 features to use in the experiment. Finally, the IP address value in the 'Address A' and 'Address B' attributes were converted to decimal. Table 4 shows a sample of the final dataset to be used in the experiment.

**Table 4**
Final dataset sample

| Label | Protocol | Address A | Port A | Address B | Port B | Packets A → B | Bytes A → B | Packets B → A | Bytes B → A |
|---|---|---|---|---|---|---|---|---|---|
| Goodware | 6 | 3232236160 | 55559 | 1177009456 | 80 | 7 | 854 | 12 | 1737 |
| Goodware | 6 | 3232237697 | 63318 | 2900408712 | 443 | 217 | 20075 | 995 | 1132562 |
| Goodware | 17 | 167772687 | 123 | 1074006561 | 123 | 1 | 90 | 1 | 90 |
| Goodware | 17 | 167772687 | 137 | 167772927 | 137 | 24 | 2640 | 24 | 2640 |
| Malware | 17 | 167772687 | 68 | 167772674 | 67 | 0 | 0 | 2 | 1180 |
| Malware | 17 | 167772687 | 123 | 884685931 | 123 | 1 | 90 | 2 | 180 |
| Malware | 6 | 167772687 | 1045 | 3638213026 | 80 | 10 | 835 | 19 | 1468 |
| Malware | 6 | 167772687 | 1048 | 876236452 | 80 | 5 | 335 | 9 | 990 |

The WEKA machine learning tool has an option to allocate a percentage split of a dataset for training and test purposes. We chose to manually split our dataset into training and test datasets to ensure each dataset contained records relating to an equal number of malware and goodware samples. Due to the difference in the number of conversations created and subsequently extracted for each sample, the number of instances in each dataset is different; the training dataset contained 75,618 instances and the test dataset contained 48,526 instances. This equates to a percentage split of 60.91% for training, and 39.09% for testing of the NetConverse model. Each dataset was finally converted into Attribute-Relation File Format (.ARFF) for processing within WEKA.

3.3 Machine Learning classifiers

It is during this stage of the experiment that we identified the machine learning classifier and feature combination that achieved the highest detection rate. Table 5 outlines the 6 machine learning classifiers that we used.

**Table 5**
Machine Learning Classifiers [34]

| Classifier | Pros | Cons |
|---|---|---|
| Bayes Network | Fast computation and data training | Impractical for large featured datasets |
| Multilayer Perceptron | Accurate estimation | High processing time |
| J48 | Fast and scalable classifier of decision tree | Can be less effective on predicting the value of continuous class attributes |
| K Nearest Neighbours (IBK) | Simple, requires no training | High processing time |
| Random Forest | Can improve predictive performance | Output can be hard to analyse |
| Logistic Model Tree (LMT) | Flexible and accurate | Potentially high bias |

The experiments were performed within a virtual VMWare workstation environment running on Kali Linux 2017.1 and Debian 4.9.25 OS; 4GB memory was allocated from the host. The host laptop was a MacBook Pro with a 2.9Ghz Intel i7 processer, 16GB DDR3 RAM and MacOS Sierra 10.12.4 OS. The machine learning tool used was WEKA 3.8.1 with a Java Runtime Version of 1.8.0_131. In the experiment, all 6 classifiers used their default values.

## 4. Experiments and Results

In this section, we present the results of our experiment and evaluate the classifiers we have used to achieve the best detection rate.

The experiment was split into 2 phases; the first phase ran 10-fold cross-validation using all 10 extracted attributes; the second ran 10-fold cross-validation using 8 extracted attributes (the Packets A → B and Packets B → A attributes were removed at this stage). In both phases each classifier model was re-evaluated against the supplied test dataset to evaluate its effectiveness.

4.1 Evaluation measures

We evaluated the performance using the 5 standard WEKA metrics: true positive rate (TPR), false positive rate (FPR), precision, recall, F-measure. The metrics are summarised in Table 6.

**Table 6**
Evaluation metrics

| Metric | Calculation | Value |
| --- | --- | --- |
| True positive rate (TPR) | TP/(TP+FN) | Correct classification of predicted malware |
| False positive rate (FPR) | FP/(FP+TN) | Goodware incorrectly predicted as malware |
| Precision | TP/(TP+FP) | Rate of relevant results |
| Recall | TPR | Sensitivity for the most relevant results |
| F-measure | 2 × (Recall × Precision)/(Recall + Precision) | Estimate of entire system performance |

TP= True positive, FN=false negative, TN= true negative, FP=false positive

4.2 Latest malware experiment and results

Table 7 summaries the time taken to build each model in each phase of the experiment (with/without feature selection). Only 2 classifiers (KNN and LMT) experienced an increase in time taken to build the model; the other 4 classifiers (Bayes Network, Multilayer Perceptron, J48 and Random Forest) experienced a decrease in processing time due to a reduction in the number of attributes to be processed.

**Table 7**
Comparison of processing time (in seconds)

| Classifier | Without Feature Selection (10 attributes) | Feature Selection (8 attributes) |
| --- | --- | --- |
| Bayes Network | 0.72 | 0.59 |
| Multilayer Perceptron | 48.9 | 36.99 |
| J48 | 0.95 | 0.18 |
| KNN | 0.01 | 0.03 |
| Random Forest | 5.46 | 4.75 |
| LMT | 28.5 | 32.7 |

Table 8 lists the results of our experiment without feature selection (10 attributes) and with feature selection (8 attributes). Only the random forest classifier demonstrated slightly decreased results with a higher FPR (+0.40%) and lower TPR, precision, recall and f-measure values. The Bayes Network and Multilayer Perceptron classifiers demonstrated a very small increase overall (0.10% and 0.30% TPR increase respectively). The KNN and LMT classifier results did not change.

**Table 8**
Comparison without and with feature selection

|  | Without Feature Selection | | | | |
| --- | --- | --- | --- | --- | --- |
| Classifier | TPR (%) | FPR (%) | Precision | Recall | F-measure |
| Bayes Network | 94.90 | 5.10 | 95.00 | 94.90 | 94.90 |
| Multilayer Perceptron | 94.90 | 6.50 | 94.80 | 94.90 | 94.90 |
| J48 | 97.10 | 1.60 | 97.30 | 97.10 | 97.10 |
| KNN | 95.30 | 4.10 | 95.50 | 95.30 | 95.30 |
| Random Forest | 96.10 | 3.70 | 96.20 | 96.10 | 96.10 |
| LMT | 96.80 | 3.90 | 96.80 | 96.80 | 96.80 |
|  | With Feature Selection | | | | |
| Classifier | TPR (%) | FPR (%) | Precision | Recall | F-measure |
| Bayes Network | 95.00 | 4.70 | 95.10 | 95.00 | 95.00 |
| Multilayer Perceptron | 95.20 | 5.50 | 95.20 | 95.20 | 95.20 |
| J48 | 97.10 | 1.60 | 97.30 | 97.10 | 97.10 |
| KNN | 95.30 | 4.20 | 95.50 | 95.30 | 95.30 |
| Random Forest | 95.10 | 4.10 | 95.40 | 95.10 | 95.20 |
| LMT | 96.80 | 3.90 | 96.80 | 96.80 | 96.80 |

Table 9 summarises the highest performance achieved for each classifier. The results for each performance metric are shown: TPR, FPR, Precision, Recall, F-Measure. The J48 classifier achieved the best performance across all 5 metrics, with a very low false positive rate (FPR) at 1.60% being an important factor in achieving a high overall system performance (f-measure) value. The J48 classifier achieved the highest accuracy with 97.10% followed by LMT with 96.80%, Random Forest with 96.10%, KNN with 95.30%, Multilayer Perceptron with 95.20% and Bayes Network with 95.00%.

**Table 9**
Malware detection evaluation result

| Classifier | TPR (%) | FPR (%) | Precision | Recall | F-measure | Feature selection |
| --- | --- | --- | --- | --- | --- | --- |
| Bayes Network | 95.00 | 4.70 | 95.10 | 95.00 | 95.00 | With |
| Multilayer Perceptron | 95.20 | 5.50 | 95.20 | 95.20 | 95.20 | With |
| J48 | 97.10 | 1.60 | 97.30 | 97.10 | 97.10 | With/Without |
| KNN | 95.30 | 4.10 | 95.50 | 95.30 | 95.30 | Without |
| Random Forest | 96.10 | 3.70 | 96.20 | 96.10 | 96.10 | Without |
| LMT | 96.80 | 3.90 | 96.80 | 96.80 | 96.80 | With/Without |

4.3 Result comparison

As far as the authors are aware there is no directly comparable study, however there are similar machine learning techniques and approaches that we can use to substantiate our results.

In [11] Android malware was analysed using a dataset created by extracting similar, but more statistical network traffic features. This technique achieved a 99.9% TPR with the random forest classifier, in comparison to a value of 97.1% achieved using the J48 classifier in NetConverse. In [10] Windows ransomware was analysed using features obtained via dynamic analysis. A slightly lower TPR of 96.3% was achieved. The final study [35] adopted a statistical network conversation approach to analyse botnet traffic, achieving an average TPR of 95.0% in the detection of 4 different botnet applications.

**Table 10**
Comparison of results from similar studies

| Study | Description | Method | Accuracy (TPR %) |
| --- | --- | --- | --- |
| Mobile malware detection [11] | Android malware | Network conversations | 99.99 |
| NetConverse | Windows ransomware | Network conversations | 97.1 |
| EldeRan [10] | Windows ransomware | Dynamic | 96.3 |
| Peershark [35] | P2P Botnets | Network conversations | 95.0 (average) |

## 5. Conclusion and Future Works

The NetConverse model demonstrated an evaluation of different machine learning classifiers to detect Windows ransomware by analysing network traffic conversation data with a high rate of accuracy. Selected classifiers were Bayes network (BN), Decision Tree (J48), K-Nearest Neighbours (IBK), Multi-Layer Perceptron, Random Forest and Logistic Model Tree.

We trained the NetConverse model with a set of extracted network traffic features for further evaluation, using a set of different classifiers. In addition, we identified the best classifier based on the TPR value. The importance of this paper relies on the method used for collecting the samples and filtering the network conversations to remove any duplication and non-relevant attributes from our training dataset. Our experiment results (classifiers) show high performance attached with a high accuracy result.

The experiment results show a 97.1% detection rate accuracy with the decision tree (J48) classifier and 96.8% detection rate accuracy with the LMT classifier. The experiment proves that machine learning classifiers can detect ransomware based on the network traffic behaviour.

This work is a baseline for future research in which researchers can extend and develop a dataset to include other ransomware families, and enhance the detection process by extracting additional attributes. There are several areas of research that could be undertaken, for example: developing real-time ransomware detection using cloud-based machine learning classifiers and outputting detection results for use within other tools.

**Acknowledgments**

We should acknowledge and thank Virus Total for graciously providing us with a private API key for use during our research to prepare the dataset. The authors would like to thank Mr. Ali Feizollah for his assistance with the feature extraction process. This work is partially supported by the European Council 268 International Incoming Fellowship (FP7-PEOPLE-2013-IIF) grant.

10[7] "Ransomware Protection - RansomFree by Cybereason." [Online]. Available: https://ransomfree.cybereason.com/. [Accessed: 31-Mar-2017].
[8] "Darktrace | Technology." [Online]. Available: https://www.darktrace.com/technology/#machine-learning. [Accessed: 31-Mar-2017].
[9] "Cerber Ransomware Now Evades Machine Learning." [Online]. Available: http://www.darkreading.com/vulnerabilities---threats/cerber-ransomware-now-evades-machine-learning/d/d-id/1328506?_mc=NL_DR_EDT_DR_daily_20170330&cid=NL_DR_EDT_DR_daily_20170330&elqTrackId=67749c8bfb29467b8ea4140c8f2f3c25&elq=4474dd7c0c1b440ab1717aa1969e0. [Accessed: 31-Mar-2017].
[10] D. Sgandurra, L. Muñoz-González, R. Mohsen, and E. C. Lupu, "Automated Dynamic Analysis of Ransomware: Benefits, Limitations and use for Detection," no. September, 2016.
[11] F. A. Narudin, A. Feizollah, N. B. Anuar, and A. Gani, "Evaluation of machine learning classifiers for mobile malware detection," *Soft Comput.*, vol. 20, no. 1, pp. 343–357, 2016.
[12] Symantec, "The evolution of ransomware," 2015.
[13] "tshark\ -\ The\ Wireshark\ Network\ Analyzer\ 2.0.0," 2017. [Online]. Available: https://www.wireshark.org/docs/man-pages/tshark.html. [Accessed: 29-May-2017].
[14] M. Hopkins and A. Dehghantanha, "Exploit Kits: The production line of the Cybercrime economy?," in *2015 2nd International Conference on Information Security and Cyber Forensics, InfoSec 2015*, 2016, pp. 23–27.
[15] A. Feizollah, N. B. Anuar, R. Salleh, and A. W. A. Wahab, "A review on feature selection in mobile malware detection," *Digit. Investig.*, vol. 13, no. March, pp. 22–37, 2015.
[16] M. Damshenas, A. Dehghantanha, and R. Mahmoud, "A Survey on Malware propagation, analysis and detection," *Int. J. Cyber-Security Digit. Forensics*, vol. 2, no. 4, pp. 10–29, 2013.
[17] N. Milosevic, A. Dehghantanha, and K. K. R. Choo, "Machine learning aided Android malware classification," *Computers and Electrical Engineering*, 2016.
[18] M. Damshenas, A. Dehghantanha, K.-K. R. Choo, and R. Mahmud, "M0Droid: An Android Behavioral-Based Malware Detection Model," *J. Inf. Priv. Secur.*, vol. 11, no. 3, pp. 141–157, Jul. 2015.
[19] K. K. R. Azmoodeh, Amin; Dehghantanha, Ali; Conti, Mauro; Choo, "Detecting Crypto Ransomware in IoT Networks Based On Energy Consumption Footprint," *J. Ambient Intell. Humaniz. Comput.*, 2017.
[20] F. Mercaldo, V. Nardone, and A. Santone, "Ransomware Inside Out," 2016.
[21] K. Liao, Z. Zhao, A. Doupe, and G.-J. Ahn, "Behind closed doors: measurement and analysis of CryptoLocker ransoms in Bitcoin," in *2016 APWG Symposium on Electronic Crime Research (eCrime)*, 2016, pp. 1–13.
[22] D. D. Hosfelt, "Automated detection and classification of cryptographic algorithms in binary programs through machine learning," 2015.
[23] S. Ranshous, S. Shen, D. Koutra, C. Faloutsos, and N. F. Samatova, "Anomaly Detection in Dynamic Networks: A Survey," 2014.
[24] Z. Wang, X. Jiang, W. Cui, X. Wang, and M. Grace, "ReFormat: Automatic Reverse Engineering of Encrypted Messages," Springer, Berlin, Heidelberg, 2009, pp. 200–215.
[25] A. Kharaz, S. Arshad, C. Mulliner, W. Robertson, and E. Kirda, "UNVEIL: A Large-Scale, Automated Approach to Detecting Ransomware," *Usenix Secur.*, pp. 757–772, 2016.
[26] K. Cabaj, P. Gawkowski, K. Grochowski, and D. Osojca, "Network activity analysis of CryptoWall ransomware," pp. 91–11, 2015.
[27] "Weka 3 - Data Mining with Open Source Machine Learning Software in Java." [Online]. Available: http://www.cs.waikato.ac.nz/ml/weka/. [Accessed: 31-Mar-2017].